\documentclass[twocolumn,showpacs,preprintnumbers,amsmath,amssymb]{revtex4}

\usepackage{graphicx}
\usepackage{dcolumn}
\usepackage{bm}

\begin{document}


\title{Elementary Particles under the Lens of the Black Holes}

\author{George E. A. Matsas}
\email{matsas@ift.unesp.br}
\affiliation{Instituto de F\'\i sica Te\'orica, 
                Universidade Estadual Paulista\\
                Rua Pamplona 145, 01405-900, S\~ao Paulo, S\~ao Paulo
               }

\date{\today}

\begin{abstract}
After a brief review of the historical development and {\em CLASSICAL}
properties of the BLACK HOLES, we discuss how our present knowledge of 
some of their {\em QUANTUM} properties shed light
on the very concept of ELEMENTARY PARTICLE. As an illustration, we
discuss in this context the decay of accelerated protons,
which may be also relevant to astrophysics.
\end{abstract}



\maketitle

\section{
\label{sec:history}
Black Holes: Historical developments 
         }

The first black hole solution was found in 1916 by the German
astrophysicist Karl Schwarzschild few months after General 
Relativity was formulated (and little time before his 
death in the Russian front). It was a static and spherically
symmetric solution of the vacuum Einstein Eqs. described
by the line element
$$
ds^2 = (1-2 M/r) dt^2 - (1-2 M/r)^{-1} dr^2 -  r^2 d\Omega\;,
$$
where $M$ is the black hole mass. Notwithstanding it took many 
decades before the scientific community accepted that black holes 
were physical solutions which could be indeed realized in nature.  
In 1939 we can still find A. Einstein stating in the conclusions 
of an article~\cite{E}: ``the Schwarzschild singularities
do not exist in the physical reality''. This was not what 
J. Oppenheimer and his student, H. Snyder, concluded in the same
year~\cite{OS}, however, after analyzing the collapse of massive stars.

In 1938, J. Oppenheimer and G. Volkoff found that neutron stars 
had a limit for its mass beyond which they should collapse~\cite{OV}. 
In the year after, Oppenheimer and Snyder decided to analyze it in 
more detail. For technical reasons, they assumed some simplifications:
spherical symmetry, constant density, no rotation and no shock waves 
with emission of matter or radiation. Under these conditions they 
concluded that the collapse would lead eventually to a black hole 
indeed, but there remained some unclear features to be understood. 
In contrast to the description made by observers at rest on the surface
of the star who would witness a continuous collapse towards the singularity, 
asymptotic observers would see the star surface like ``frozen'' on the 
event horizon. These seemingly contradictory descriptions were only 
reconciled after D. Finkelstein found in 1958 a coordinate system 
which was able to cover smoothly the 
internal and external regions of the  black hole~\cite{F}. This conceptual step 
in addition with more precise numerical simulations, which were 
possible thanks to a better comprehension of the nuclear structure, 
ended up to corroborate Oppenheimer and Snyder's conclusion and 
bit most skepticism about the possible existence of black holes. J. Wheeler, 
in particular, evolved from criticizer to supporter of the black hole idea  
and in 1967 he introduced the denomination {\em black hole} to what was 
called {\em collapsed star} in the west and {\em frozen star} in the east. 
More than 40 years after the Schwarzschild solution was discovered, 
black holes were treated at least as a real possibility. 

It is common to consider 1964 as the beginning of the {\em black hole golden 
era}. From the theoretical point of view, R. Penrose has introduced 
topological methods with which he was able to derive some quite general 
results. For example, he was able to show (under some natural assumptions in 
the classical realm) that black holes must have a singularity in their 
interior~\cite{P}. Developments in the observational
domain also took place. In 1966 I. Novikov and Ya. Zel'dovich  raised the 
possibility that there should exist binary systems formed by an ordinary
star and a black hole orbiting around each other. It would be natural, thus, 
to expect the combined emission of X-ray and visible light from such systems 
since as matter is attracted by the black hole its gravitational 
potential would be converted into thermal energy and eventually into 
X-ray~\cite{NZ}. 
This turned out to be the most probable explanation for the spectrum
associated with Cygnus X-1 as it became 
clear in 1971 with the collected  data from the Uhuru satellite. It is 
worthwhile to notice that while the prediction  that star-size black holes 
could be X-ray sources was confirmed only 5 years after its formulation, 
the explanation that radio galaxies (observed since the 30's) and 
quasars (observed since the 60's) were energized by the presence of 
super black holes had to wait more than 40 years. 

Different evidences favoring the existence of black holes are mounting 
since then, and it is expected for soon some  direct 
signal from a black hole event horizon. This may come as a 
shadow disc at the photograph 
plate of Sgr A$^*$ (at the center of the Milky Way, where a 
many-million-solar-mass black hole is believed to exist) [when small enough 
wavelength observations become possible] or in the form of gravitational 
wave signals to be detected up to the 10's by the Ligo and Virgo Earth-based
gravitational  wave detectors or in the 20's  by the Lisa space gravitational 
wave detector, we do not know; but what we do know is
that {\em it will be the confirmation of one of the greatest predictions of 
theoretical physics}.

\section{
\label{sec:classical}
Black Holes: CLASSICAL properties
      }     

A strongly asymptotically predictable spacetime 
$$
({\cal M}, g)
$$ 
is formally said to contain a black hole $B$ if 
$$  
B \equiv {\cal M} - J^- ({\cal J^+})
$$
is not empty, i.e., if there is a region from where classical light rays
cannot escape to infinity, where $J^-$ is the causal past and ${\cal J}^+ $ 
is the future  null infinity, .
The event horizon of the black hole is defined as being the boundary of $B$:
$$
H \equiv \dot J^- ({\cal J}^+) \cap {\cal M}\;.
$$
The solution discovered by Schwarzschild contains a particular kind of
black hole which is static and spherically symmetric  
but could it exist other black holes with, let us say, more exotic forms 
and exquisite properties? 
In 1964, A. Doroshkevich, I. Novikov and Ya. Zel'dovich showed that 
quasi-spherically symmetric collapsing stars give rise to perfectly 
spherically symmetric black holes~\cite{DZN}. This was the prelude of a series 
of far-reaching theorems known as {\em black hole no-hair theorems}. 

In 1967 W. Israel derived what can be considered the first piece of 
this series of theorems, namely, {\em every rotationless black hole 
should be spherically symmetric}~\cite{I}. 
As a next step, it was natural, thus, to extend the analysis to 
rotating black holes. A solution for a rotating black hole was unveiled
by R. Kerr in 1963~\cite{K} (but only identified as so in 1965 by R. Boyer
and R. Lindquist~\cite{BL}, B. Carter~\cite{C} and R. Penrose). At that time, 
it was not clear, 
however, if there would not exist other vacuum solutions of the Einstein 
Eqs. describing black holes with angular momentum. This quest was embraced 
by B. Carter in 1972 (with a contribution by D. Robinson) who 
showed that according to the vacuum Einstein Eqs. the most general
black hole solution was the one given by Kerr. 
The event horizon of a Kerr black hole is more elongated  at the equator
than on the poles and the underlying geometry of a rotating black hole is richer 
than of a static one but still its structure remains quite simple since most 
properties of the original star are lost in the collapse. To put it in 
R. Price's words: In a star collapse process with  a black hole formation, 
everything that can be radiated (i.e. does not satisfy some 
conservation law) will be radiated. 

The most general formulation of the no-hair theorems associated
with the electrovacuum solution of Einstein Eqs. states that 
black holes are completely characterized by their mass $M$, charge $Q$ 
and angular momentum $J$ and its geometry is described by the Kerr-Newman 
line element. For instance, the black hole area can be written as ($c=G=1$)
$$
A = 4\pi^2 \left[ 2 M^2 - Q^2 + 2 M \sqrt{M^2 - Q^2 - J^2/M^2\,} \right]\;.
$$

Thus black holes not only are probably the most
exotic structures at the heavens but also one of the simplest ones
as well. 

\section{
\label{sec:semiclassical}
Black Holes: SEMICLASSICAL properties
      } 
      
The beginning of the black hole semiclassical era took place in 1974.
This was the summit of a number of curious events which actually began
in 1971~\cite{H71}. In this year, S. Hawking showed that the total
horizon area for any given set of black holes did not decrease 
with time. In particular, according to this theorem, black holes
were indestructible. In order to derive this theorem, Hawking used
some quite reasonable hypotheses (at least in the classical realm).
In 1972, in analogy to the second law of thermodynamics, J. Bekenstein 
associated an entropy to each black hole proportional to the area
of its event horizon~\cite{B}. Hawking had a strong negative reaction at
first but two years later, as he analyzed the collapse of stars in the context 
of Quantum Field Theory in Curved Spacetimes (where positive energy conditions
normally used in classical theorems are not valid),  Hawking 
showed that black holes should radiate with a thermal spectrum~\cite{H74} 
with temperature ($c=G=\hbar=k_B=1$)
$$
T = {\cal K}/ 2\pi\;,
$$
(as measured by assymptotic observers), where 
$$
{\cal K} = 4 \pi \sqrt{M^2 - Q^2 - J^2/M^2\,}/A
$$
is the surface gravity.  Eventually, black holes could have associated an 
entropy proportional to its horizon area  
$$
S = \frac{c^3 A}{4G \hbar}
$$
as conjectured by Bekenstein (and precisely calculated by Hawking).
This discovery opened a subarea denominated {\em Black Hole Thermodynamics},
which is presently very active because of some fundamental questions
raised in connection with information theory and quantum mechanics but
which will be hardly solved outside the context of a full quantum gravity
theory. In Hawking´s words: {\em Holes may be black classically but are  
gray quantum-mechanically}. 

In order to understand better the Hawking effect, let us make a 
detour through Quantum Field Theory.  It became clear since the early 
times of Quantum Mechanics   
that the no-particle state, i.e. the vacuum, has a very rich structure.
Most (if not all) of its exotic  properties are connected with the 
concept of virtual particles. Virtual particles violate the 
Heisenberg uncertainty principle and, thus, cannot be directly 
observed. Notwithstanding, they do have indirect observable 
consequences. Probably the most paradigmatic example of the
physical consequences of the virtual particles is given by the
Casimir effect.

According to the Casimir effect~\cite{Ca}, uncharged parallel metallic 
plates in the vacuum experience an attractive pressure given
by (see Ref.~\cite{MT} for a comprehensive review and Ref.~\cite{C-PFT}
for a pedagogical introduction)
$$
|F|/A = 3 \pi^2  \hbar c/ 710 \, d^4 \;,
$$ 
where $d$ is the distance between
the plates and we are discounting any 
gravitational effects because of the plate masses. We note 
that this is  intrinsically a quantum-relativistic effect which 
would vanish for $\hbar \to 0$ and lead to nonsense results
in the nonrelativistic limit $c \to \infty$. Roughly speaking,
the metal plates play the role of boundaries to the virtual
photons diminishing the total vacuum energy  
$\langle 0 | \hat H | 0 \rangle$  as the plates get closer to 
each other, where $\hat H$ is the free Hamiltonian associated 
with the photon field.

We already know that virtual photons feel the presence of
static metallic plates but what does it happen if we consider a 
(nonuniformly) accelerated metallic plate in the vacuum? 
The metal plate will transfer energy to the virtual particles
letting them real. Indeed, a photon flux will be emitted
opposite to the acceleration direction while negative
energy fluxes will be emitted in the acceleration direction.
This is known as dynamical Casimir effect (but could be fairly called 
Moore effect~\cite{M}). 
This effect is interesting in its own right and also for being a kind of 
flat-spacetime analog of the Hawking effect. Here the mirror plays 
the role of the star, the emitted photons correspond to the 
Hawking radiation and the inward flux of negative energy  
is responsible for the black hole evaporation. The main difference
here is that contrary to the mirror case, where only photons
are radiated, the star collapse leads to the emission of
all kind of particles. This is so because, according to the 
equivalence principle, all particles are coupled to gravity 
in the same way. What would not be easy to anticipate 
is that the spectrum of the emitted particles as detected by
asymptotic observers can be associated to a black body. In the particular
case of a static chargeless black hole, the corresponding temperature
is
$$
T= \hbar c^3 /  8 \pi k_B G M\;,
$$ 
where $M$ is the black hole mass. Notice the appearance of the four 
universal constants $c,\hbar,G, k_B$.

The larger the black hole, the 
lower the temperature and only ``small-mass'' particles 
($m c^2 \leq k_B T$) will be likely to escape. 
Large-mass particles will be scattered back to the hole by the
scattering potential. Notwithstanding, it is worthwhile to notice that
arbitrarily large mass particles could be, in principle, observed as 
follows. By assuming that the evaporation process is adiabatic, 
the radiation temperature as measured 
by static observers at different Schwarzschild radial coordinates 
$r$  outside the black hole  will differ from the one at the infinity 
by a red-shift factor~\cite{T}, namely, 
$$
T(r)= T / \sqrt{1-2 GM/r c^2}\;.
$$
Thus, the closer to the horizon the higher the temperature and the
more likely to detect massive particles. However, there is no free
lunch in nature: in order to probe particles with Planck mass one 
has to get as close to the horizon  as the 
Planck length.  

\section{
\label{sec:elementaryparticles}
ELEMENTARY PARTICLES UNDER THE LENS OF THE BLACK HOLES
      } 
      
The Hawking effect connects in a nontrivial way Relativity, Quantum 
Mechanics, Gravity and Thermodynamics and has raised a number of different
questions, some of which are still opened. Notwithstanding it became
clear since 1976 after W. Unruh~\cite{U} that although static observers outside 
black holes detect a thermal bath of particles, free falling observers 
close enough to the horizon would have their detectors basically 
unexcited. (Here one may think of a usual 2-level Unruh-DeWitt 
detector~\cite{UD}.) 
The explanation for this phenomenon is closely connected with 
previous works by S. Fulling~\cite{Fu}
and P. Davies~\cite{D} which called attention to the fact that the particle 
content 
of a Quantum Field Theory  is observer dependent. This conclusion has
far-reaching implications even to Quantum Field Theory in flat spacetime. 
Indeed, the vacuum state  as defined by inertial observers in the 
Minkowski space  corresponds to a thermal state of all particles at 
temperature
$$
T = \hbar a/2 \pi c k_B
$$ 
as detected by observers with constant
proper acceleration $a$. It can be said that uniformly accelerated observers 
see
as real those particles which inertial observers ascribe as being virtual.
 
It is also possible to figure out the opposite situation where 
particles which are unobservable to uniformly accelerated observers are 
observable to inertial ones. In 1991 A. Higuchi, D. Sudarsky and the author 
were analyzing
the following problem associated with the radiation emitted from uniformly 
accelerated charges. It is well known that accelerated charges
radiate with respect to inertial observers and the emitted power is given 
by the
Larmor formula~\cite{J} as (see also Ref.~\cite{H} for a deep discussion on 
the radiation reaction problem)
$$
W = e^2 a^2 /6 \pi c^3 \;.
$$  

In spite of this, there was a 
consensus that co-accelerated observers with  uniformly accelerated charges, 
i.e. charges with constant proper acceleration $a$, would not detect any 
radiation since the corresponding field is static with respect to 
them~\cite{RB}. According to Quantum Field 
Theory, however, the usual classical electromagnetic radiation can be 
interpreted  in terms of photons. So, if the co-accelerated observers did not
observe any radiation, ``where had the photons observed by the inertial 
observers gone"? 
The answer to this question is directly related with the fact that the 
elementary particle
concept is observer dependent. Indeed, the emission of a finite-energy
photon as seen in the inertial frame corresponds to the emission to or 
absorption from  the Fulling-Davies-Unruh (FDU) thermal bath (in which the 
electron is immersed according to co-accelerated observers) of a 
{\em zero-energy} 
Rindler photon. The emission rate of finite energy photons as defined by the 
inertial observers and the combined emission and absorption rate of 
zero-energy 
Rindler photons as defined by the co-accelerated observers can be both written 
as~\cite{HMS1} ($c, \hbar = 1$)
$$
P_{k_\bot} (a) = \frac{e^2}{4 \pi^3 a} | K_1 (k_\bot/a)|^2
$$
where $k_\bot$ is the photon transverse momentum (with respect to the 
acceleration 
direction). 
Zero-energy Rindler photons are perfectly well defined
entities since they can carry non-zero transverse momentum 
but cannot be detected by physical observers
because they concentrate on the horizon of the uniformly accelerated 
observers~\cite{CCMV}.
From an epistemological point of view, zero-energy Rindler photons
have much in common with virtual particles since although they cannot 
be observed
they are indirectly important as a mean to explain some physical phenomena;
in this case, the ``disappearance" of the photons in the electron 
co-accelerated frame. Zero-energy particles are also important in analyzing
other problems as, for instance, the response of static sources 
interacting with the 
Hawking radiation of a black hole~\cite{CCHMS}. 

Probably because of its non-intuitiveness  the FDU 
effect was received with skepticism by part of the scientific 
community. Although the derivation of the effect is sound and the
conclusion indisputable, part of the community took  the position  
that only a ``{\em direct}" observation of the effect would be convincing.
Notwithstanding, this is not an easy task since no macroscopic 
body would resist to the typical accelerations $a$ necessary 
for this purpose:
$$
T/ (1 K) = a/ (2.5 \times 10^{22} cm/s^2)\;.
$$
The strategy had to be otherwise, namely, a gedanken experiment able
to make it clear that the FDU effect would be necessary
for the consistency of the Quantum Field Theory itself. This
 was the strategy
followed by D. Vanzella and the author~\cite{MV} inspired by  previous 
works~\cite{UW,HMS1}.

According to the standard model, inertial protons are stable. But this is
not so for accelerated ones because of the work transferred to the proton
by the external accelerating agent. As far as the proton proper acceleration 
satisfies $a \ll m_n + m_e + m_\nu - m_p$ the decay process will 
be much suppressed
but for  $a > m_n + m_e + m_\nu - m_p$ the weak decay channel
\begin{equation}
p^+ \to n^0 + e^+ + \nu
\label{pweakdecay}
\end{equation}
will be favored up to $a \approx m_\pi $ after what the strong-decay channel
\begin{equation}
p^+ \to n^0 + \pi^+
\label{pstrongdecay}
\end{equation}
will dominate. Recent calculations show that high-energy protons with 
$E\approx 10^{14}$ eV under the influence of magnetic fields of 
$B \approx 10^{14}$ G found in some pulsars should decay in a fraction of 
a second in laboratory time~\cite{VM}. 

The analysis above, however, is valid for inertial observers. But how
can we understand the decay process from the point of view of co-moving
observers with a uniformly accelerated proton? According to these observers, 
in order to decay the proton 
must remove energy from the particles of the thermal bath
in which it is immersed in its rest frame. Thus, according to the co-moving
observers, the decay processes  will be
seen quite differently. Indeed, in the regime where the proton/neutron can be 
considered as  unexcited/excited states of a two-level quantum mechanical
system, processes (\ref{pweakdecay}) and (\ref{pstrongdecay}) will be 
interpreted 
according to coaccelerated observers as
\begin{subequations}
\label{pweakdecayac}
\begin{equation}
p^+ + e^- \to  n^0  + \nu
\end{equation}
\begin{equation}
p^+ + \bar \nu  \to  n^0 + e^+ 
\end{equation}
\begin{equation}
p^+ + \bar \nu + e^-  \to   n^0 
\end{equation}
\end{subequations}
and
\begin{equation}
p^+ + \pi^- \to n^0 
\label{pstrongdecayac}
\end{equation}   
respectively. In particular, the correct mean lifetime  is predicted in the 
co-accelerated frame by assuming the processes above in conjunction with 
the presence of the FDU thermal bath~\cite{MV,SY}. Had we not taken 
into account 
the FDU thermal
bath, the proton would be seemingly stable according to the co-accelerated 
observers 
(for sake  of energy conservation) in contradiction with the inertial frame 
conclusion: {\em The FDU effect is necessary for the consistency of 
Quantum Field 
Theory.}
    
\section{
\label{sec:concludingthought}
Concluding remarks      } 

The overwhelming difficulty of constructing a quantum gravity theory can
be illustrated  by the fact that different people will give different answers
to what such a theory should look like. Moreover, there is no reason to 
believe that the rules of quantum mechanics which are tested up to scales
of $10^{-15}$ cm would not be drastically modified in the quantum gravity 
domain. Assuming that $c$, $\hbar$, and $G$ are the only fundamental constants
to the quantum gravity theory, we expect that its typical effects will become 
obvious at the Planck scale, i.e. as soon as we accelerate elementary 
particles at energies of $E > M_p c^2 = \sqrt{\hbar c^5/G} $, we are 
able to probe distances of $L_p < \sqrt{G \hbar/c^3}$, we look at 
processes with
time scales of  $T_p < \sqrt{G \hbar/c^5}$ or we observe structures
with densities of $\rho_p = c^5/G^2 c^2$. In principle, these extreme 
situations would be likely to be realized only in singular regions, 
as e.g.,  
close to the Big Bang 
and at black hole singularities. Unhappily the Big Bang is mostly screened by
a number of effects associated with the primordial plasma (although it may
be that gravitational wave detectors open a window to it) and black hole
singularities are not naked. Thus, it might seem that we would be hopelessly
lost from both sides: theoretically and observationally. But this is not so 
according to the Semiclassical Gravity theory! 

If one is not allowed to visit the 
Chinese Imperial city, one should better wait for news just outside its limits. 
In our case, the Imperial city is the Quantum Gravity realm; it is forbidden
to us because we do not fit into the Planck scale; and it is worthwhile to 
wait for news coming from it because  quantum mechanical information 
should leak 
from the lock.
The Hawking effect is probably the better example of how quantum gravity 
effects
can escape towards the macroscopic domain. It might 
be difficult to observe the radiation emitted from large black holes since 
the associated temperature is very small:
$$
T/(1 K) = 10^{-7} M_\odot / M
$$  
but this is not so for the radiation emitted from smaller (primordial?) 
black holes. 
Even for large black holes, the situation is not that bad as soon as we may
probe directly the region close to the horizon 
where the radiation temperature is very blue-shifted. 

We do not know how far we will be 
able to go with this semiclassical approach as well as people did not know how 
far they were going 
to reach by using the semiclassical electromagnetic theory rather than 
QED in atomic 
physics; but what we do know is that every step forward in this down-up 
strategy will
be (in principle) a long-lasting one because, after all, we are dealing 
with the safe 
side of our standard theories. Moreover because the Semiclassical Gravity 
is in the 
interface of General Relativity, Quantum Field Theory, and Thermodynamics, 
unexpected
effects which does not have to do directly with Quantum Gravity are being 
unveiled. 
Here we have focused on the contribution of the Semiclassical Gravity 
Theory to the 
concept of elementary particle
but other contributions could also be 
cited. Recently, Unruh has raised the very interesting possibility of 
mimicking 
the Hawking 
effect through Condensed Matter laboratory experiments~\cite{U2}. 
For this purpose 
it is enough to arrange a compact region in a background medium (think of a
spherical region in the middle of a pool) such that inside it the  
inward velocity 
of the medium is larger than the  sound velocity. In this way, phonons 
would not 
be able to escape from this trapped  region and we would have a 
sonic hole. 
Many (kinematical) classical and semiclassical properties of the 
black holes can be 
experimentally probed in this way. In particlular, Hawking phonon radiation 
is expected to be observed from sonic holes. 

More embarassing than having not formulated yet the full quantum
gravity theory is being aware of how much we still do not know 
about those theories which we thought to have mastered long ago. In this 
vein, quantum gravity can wait; the misteries hidden in our standard 
theories cannot. After all, we can always hold on V. Weisskopf words: 
{\em Is it really the end of theoretical physics to get the world formula? 
The greatest physicists have always thought that there was one, 
and that everything else could be derived from it. Einstein believed it, 
Heisenberg believed it. I am not such a great physicist, I do not 
believe it... This, I think, is because nature is inexhaustible.}

\begin{acknowledgments}
I am thankful to J. Casti\~neiras and A. Dias for reading the manuscript and 
Conselho Cient\'\i fico e Tecnol\'ogico and Funda\c c\~ao de Amparo \`a 
Pesquisa do Estado de S\~ao Paulo for funding partially this work and 
consistent support along the years.
\end{acknowledgments}



\end{document}